\documentclass[twocolumn,epjc3]{svjour3}
\usepackage{graphicx}
\usepackage{dcolumn}
\usepackage{bm}
\usepackage{amsmath}
\usepackage{amssymb}

\RequirePackage[T1]{fontenc}

\smartqed  

\RequirePackage{graphicx}
\RequirePackage{mathptmx}      
\RequirePackage{flushend}
\RequirePackage[colorlinks,citecolor=blue,urlcolor=blue,linkcolor=blue]{hyperref}

\journalname{Eur. Phys. J. C}

\begin{document}

\title{Chaos in two black holes with next-to-leading order spin-spin interactions
} 


\author{Guoqing Huang \thanksref{addr1}
        \and
        Xiaoting Ni \thanksref{addr1} 
 \and
        Xin Wu \thanksref{e1,addr1}
}

\thankstext{e1}{e-mail: xwu@ncu.edu.cn}

\institute{Department of Physics, Nanchang University, Nanchang
330031, China \label{addr1}
  }

\date{Received: date / Accepted: date}

\maketitle

\begin{abstract}

 We take into account the dynamics of a complete
third post-Newtonian conservative Hamiltonian of two spinning black
holes, where the orbital part arrives at the third post-Newtonian
precision level and the spin-spin part with the spin-orbit part
includes the leading-order and next-to-leading-order contributions.
It is shown through numerical simulations that the next-to-leading
order spin-spin couplings play an important role in chaos. A
dynamical sensitivity to the variation of single parameter is also
investigated. In particular, there are a number of
\textit{observable} orbits whose initial radii are large enough and
which become chaotic before coalescence.

\end{abstract}

\section{Introduction}

Massive binary black-hole systems are likely the most promising
sources for future gravitational wave detectors. The successful
detection of the waveforms means using matched-filtering techniques
to best separate a faint signal from the noise and requires a very
precise modelling of the expected waveforms. Post-Newtonian (PN)
approximations can satisfy this requirement. Up to now,
high-precision PN templates have already been  known for the
non-spin part up to 3.5PN order (\textit{i.e.} the order $1/c^{7}$
in the formal expansion in powers of $1/c^{2}$ with $c$ being the
speed of light) [1,2], the spin-orbit part up to 3.5PN order
including the leading-order (LO, 1.5PN), next-to-leading-order (NLO,
2.5PN) and next-to-next-to-leading-order (NNLO, 3.5PN) interactions
[3-5], and the spin-spin part up to 4PN order consisting of the LO
(2PN), NLO (3PN) and NNLO (4PN) couplings [6-9].

However, an extremely sensitive dependence on initial conditions as
the basic feature of chaotic systems would pose a challenge to the
implementation of such matched filters, since the number of filters
required to detect these waveforms is exponentially large with
increasing detection sensitivity. This has led to some authors
focusing on research of chaos in the orbits of two spinning black
holes. Chaos was firstly found and confirmed in the 2PN Lagrangian
approximation of comparable mass binaries with the LO spin-orbit and
LO spin-spin effects [10]. Moreover, it was reported in [11] that
the presence of chaos should be ruled out in these systems because
no positive Lyapunov exponents could be found. As an answer to this
claim, Refs. [12,13] obtained some  positive Lyapunov exponents and
pointed out these zero Lyapunov exponents of [11] due to the less
rigorous calculation of the Lyapunov exponents of two nearby orbits
with unapt rescaling. In fact, the conflicting results on Lyapunov
exponents are because the two papers [11,12] used different methods
to compute their Lyapunov exponents, as was mentioned in [14]. Ref.
[11] computed the stabilizing limit values of Lyapunov exponents,
and Ref. [12] worked out the slopes of fit lines. This is the
so-called doubt regarding different chaos indicators causing two
distinct claims on the chaotic behavior. Besides this, there was
second doubt on different dynamical approximations making the same
physical system have distinct dynamical behaviors. The 2PN
harmonic-coordinates Lagrangian formulation of the two-black hole
system with the LO spin-orbit couplings of one spinning body allows
chaos [15], but the 2PN ADM (Arnowitt-Deser-Misner) -coordinates
Hamiltonian does not [16,17]. Levin [18] thought that there is no
formal conflict between them since the two approaches are not
exactly but approximately equal, and different dynamical behaviors
between the two approximately related systems are permitted
according to dynamical systems theory. Seen from the canonical,
conjugate spin coordinates [19], the  former non-integrability and
the latter integrability are clearer. As extensions, both any PN
conservative Hamiltonian binary system with one spinning body and a
conservative Hamiltonian of two spinning bodies without the
constraint of equal mass or with the spin-orbit couplings not
restricted to the leading order are still integrable. Recently,
[20,21] argued the integrability of the 2PN Hamiltonian without the
spin-spin couplings and with the NLO and/or NNLO spin-orbit
contributions included. On the contrary, the corresponding
Lagrangian counterpart with spin effects limited to the spin-orbit
interactions up to the NLO terms exhibits the stronger chaoticity
[22]. Third doubt relates to different dependence of chaos on single
dynamical parameter or initial condition. The description of the
chaotic regions and chaotic parameter spaces in [15] are
inconsistent with that in [23]. The different claims are regarded to
be correct according to the statement of [24] that chaos does not
depend only on single physical parameter or initial condition but a
complicated combination of all parameters and initial conditions.

It is worth emphasizing that the spin-spin effects are the most
important source for causing chaos in spinning compact binaries, but
they were only restricted to the LO term in the published papers on
research of the chaotic behavior. It should be significant to
discuss the NLO spin-spin couplings included to a contribution of
chaos. For the sake of this, we shall consider a complete 3PN
conservative Hamiltonian of two spinning black holes, where the
orbital part is up to the 3PN order and the spin-spin part as well
as the spin-orbit part includes the LO and NLO interactions. In this
way, we want to know whether the inclusion of the NLO spin-spin
couplings have an effect on chaos, and whether there is chaos before
coalescence of the binaries.

\section{Third post-Newtonian order Hamiltonian approach}

It is too difficult to strictly describe the dynamics of a system of
two mass comparable spinning black holes in general relativity.
Instead, the PN approximation method is often used. Suppose that the
two bodies have masses $m_1$ and $m_2$ with $m_1\leq m_2$.  Other
mass parameters are the total mass $M=m_1+m_2$, the reduced mass
$\mu=m_{1}m_{2}/M$, the mass ratio $\beta=m_1/m_2$ and the mass
parameter $\eta=\mu/M=\beta/(1+\beta)^{2}$. As to other specified
notations, a 3-dimensional vector $\bm r$ represents the relative
position of body 1 to body 2, its unit radial vector is $\bm n=\bm
r/r$ with the radius $r=|\bm r|$, and $\bm p$ stands for the momenta
of body 1 relative to the centre. The momenta, distances and time
$t$ are respectively measured in terms of $\mu$, $M$ and $M$.
Additionally, geometric units $c=G=1$ are adopted. The two spin
vectors are $\mathbf{S}_{i}=S_{i}\mathbf{\hat{S}}_{i}$ ($i=1,2$)
with unit vectors $\mathbf{\hat{S}}_{i} $ and the spin magnitudes
$S_i=\chi_{i}m_{i}^{2}/M^2$ ($0\leq\chi_{i}\leq 1$). In ADM
coordinates, the system can be expressed as the dimensionless
conservative 3PN Hamiltonian
\begin{eqnarray}
H(\mathbf{r},\mathbf{p},\mathbf{S}_1,\mathbf{S}_2) &=&
H_{o}(\mathbf{r},\mathbf{p})+H_{so}(\mathbf{r},\mathbf{p},\mathbf{S}_1,\mathbf{S}_2)
\nonumber
\\ & & +H_{ss}(\mathbf{r},\mathbf{p},\mathbf{S}_1,\mathbf{S}_2).
\end{eqnarray}
In the following, we write its detailed expressions although they
are too long.

For the conservative case, the orbital part $H_{o}$ does not include
the dissipative 2.5PN term (which is the leading order radiation
damping level) but  the Newtonian term $H_{N}$ and the PN
contributions $H_{1PN}$, $H_{2PN}$ and $H_{3PN}$, that is,
\begin{equation}
H_{o}=H_{N}+H_{1PN}+ H_{2PN}+H_{3PN}.
\end{equation}
As given in [25], they are
\begin{eqnarray}
H_{N}&=&\frac{\bm{p}^2}{2}-\frac{1}{r},
\end{eqnarray}
\begin{eqnarray}
H_{1PN}&=&\frac{1}{8}(3\eta-1)\bm{p}^4-\frac{1}{2}[(3+\eta)\bm{p}^2+\eta(\bm{n}\cdot\bm{p})^2]\frac{1}{r}
\nonumber \\& &+\frac{1}{2r^{2}},
\end{eqnarray}
\begin{eqnarray}
H_{2PN}&=&\frac{1}{16}(1-5\eta+5\eta^2)\bm{p}^6+\frac{1}{8}[(5-20\eta-3\eta^2)\bm{p}^4
\nonumber \\& &
-2\eta^2{(\bm{n}\cdot\bm{p})^2}\bm{p}^2-3\eta^2{(\bm{n}\cdot\bm{p})^4}]
\frac{1}{r}+\frac{1}{2}[(5+8\eta)\bm{p}^2
\nonumber \\
& & +3\eta
(\bm{n}\cdot\bm{p})^2]\frac{1}{r^2}-\frac{1}{4}(1+3\eta)\frac{1}{r^3},
\end{eqnarray}
\begin{eqnarray}
H_{3PN}&=&\frac{1}{128}(-5+35\eta-70\eta^2+35\eta^3)\bm{p}^8+\frac{1}{16}[(-7
\nonumber \\
&
&+42\eta-53\eta^2-5\eta^3)\bm{p}^6+(2-3\eta)\eta^2(\bm{n}\cdot\bm{p})^2
\nonumber \\
& &
\times\bm{p}^4+3(1-\eta)\eta^2(\bm{n}\cdot\bm{p})^4\bm{p}^2-5\eta^3(\bm{n}\cdot\bm{p})^6]\frac{1}{r}
\nonumber \\
& &+[\frac{1}{16}(-27+136\eta+109\eta^2) \bm{p}^4+\frac{1}{16}(17
\nonumber \\
&
&+30\eta)\eta(\bm{n}\cdot\bm{p})^2\bm{p}^2+\frac{1}{12}(5+43\eta)\eta(\bm{n}\cdot\bm{p})^4]\frac{1}{r^2}
\nonumber \\
&
&+\{[-\frac{25}{8}+(\frac{1}{64}\pi^2-\frac{335}{48})\eta-\frac{23}{8}\eta^2]\bm{p}^2
\nonumber \\
&
&+(-\frac{85}{16}-\frac{3}{64}\pi^2-\frac{7}{4}\eta)\eta(\bm{n}\cdot\bm{p})^2\}\frac{1}{r^3}
\nonumber \\
&
&+[\frac{1}{8}+(\frac{109}{12}-\frac{21}{32}\pi^2)\eta]\frac{1}{r^4}.
\end{eqnarray}

The spin-orbit part $H_{so}$ is linear functions of the two spins.
It is the sum of the LO spin-orbit term $H^{LO}_{so}$ and the NLO
spin-orbit term $H^{NLO}_{so}$, \textit{i.e.}
\begin{equation}
H_{so}(\mathbf{r},\mathbf{p},\mathbf{S}_1,\mathbf{S}_2)=H^{LO}_{so}
(\mathbf{r},\mathbf{p},\mathbf{S}_1,\mathbf{S}_2)
+H^{NLO}_{so}(\mathbf{r},\mathbf{p},\mathbf{S}_1,\mathbf{S}_2).
\end{equation}
Ref. [5] gave their expressions
\begin{eqnarray}
H_{so}&=&\frac{1}{r^{3}}[g(\bm{r},\bm{p})\bm{S}+g^{*}(\bm{r},\bm{p})\bm{S}^*]\cdot\bm{L},
\end{eqnarray}
where the related notations are
\begin{eqnarray*}
\bm{S}=\bm{S}_{1} + \bm{S}_{2},  \bm{S}^{*}=
\frac{1}{\beta}\bm{S}_{1} + \beta\bm{S}_{2},
\end{eqnarray*}
\begin{eqnarray*}
g(\bm{r},\bm{p})&=& 2+[\frac{19}{8}\eta\mathbf{p}^{2}+\frac{3}{2}
\eta(\mathbf{n}\cdot\mathbf{p})^{2}-(6+2\eta)\frac{1}{r}],
\end{eqnarray*}
\begin{eqnarray*}
g^{*}(\mathbf{r},\mathbf{p})&=&
\frac{3}{2}+[-(\frac{5}{8}+2\eta)\mathbf{p}^{2}+\frac{3}{4}\eta(\mathbf{n}\cdot\mathbf{p})^{2}
\nonumber \\ & & -(5+2\eta)\frac{1}{r}],
\end{eqnarray*}
and the Newtonian-looking orbital angular momentum vector is
\begin{equation}
\mathbf{L}=\mathbf{r}\times\mathbf{p}.
\end{equation}
The constant terms in $g$ and $g^{*}$ correspond to the LO part, and
the others, the NLO part.

Similarly, the spin-spin Hamiltonian $H_{ss}$  also consists of the
LO spin-spin coupling term $H^{LO}_{ss}$ and the NLO spin-spin
coupling term $H^{NLO}_{ss}$, namely,
\begin{equation}
H_{ss}(\mathbf{r},\mathbf{p},\mathbf{S}_1,\mathbf{S}_2)=H^{LO}_{ss}(\mathbf{r},
\mathbf{S}_1,\mathbf{S}_2)+H^{NLO}_{ss}
(\mathbf{r},\mathbf{p},\mathbf{S}_1,\mathbf{S}_2).
\end{equation}
The first sub-Hamiltonian reads [25]
\begin{eqnarray}
H^{LO}_{ss} &=&
\frac{1}{2r^{3}}[3(\mathbf{S}_0\cdot\bm{n})^{2}-\mathbf{S}^{2}_0]
\end{eqnarray}
with $\mathbf{S}_0=\mathbf{S}+\mathbf{S}^{*}$. The second
sub-Hamiltonian is made of three parts,
\begin{equation}
H^{NLO}_{ss}=H_{s^{2}_{1}p^{2}}+H_{s^{2}_{2}p^{2}}+H_{s_{1}s_{2}p^{2}}.
\end{equation}
They are written as [7,8]
\begin{eqnarray}
H_{s^{2}_{1}p^{2}}&=&\frac{\eta^{2}}{\beta^{2}r^{3}}[\frac{1}{4}(\bm{p}_{1}\cdot\bm{S}_{1})^{2}
+\frac{3}{8}(\bm{p}_{1}\cdot\bm{n})^{2}\bm{S}^{2}_{1} \nonumber \\ &
& -\frac{3}{8}\bm{p}^{2}_{1}(\bm{S}_{1}\cdot\bm{n})^{2}-\frac{3}{4}
(\bm{p}_{1}\cdot\bm{n})(\bm{S}_{1}\cdot\bm{n})(\bm{p}_{1}\cdot\bm{S}_{1})]
\nonumber \\ & &
-\frac{\eta^{2}}{r^{3}}[\frac{3}{4}\bm{p}^{2}_{2}\bm{S}^{2}_{1}
-\frac{9}{4}\bm{p}^{2}_{2}(\bm{S}_{1}\cdot\bm{n})^{2}] \nonumber \\
& & +\frac{\eta^{2}}{r^{3}\beta}[\frac{3}{4}
(\bm{p}_{1}\cdot\bm{p}_{2})\bm{S^{2}_{1}}-\frac{9}{4}(\bm{p}_{1}\cdot\bm{p}_{2})(\bm{S}_{1}\cdot\bm{n})^{2}
\nonumber \\ & &
-\frac{3}{2}(\bm{p}_{1}\cdot\bm{n})(\bm{p}_{2}\cdot\bm{S}_{1})(\bm{S}_{1}\cdot\bm{n})\nonumber
\\ & &
+3(\bm{p}_{2}\cdot\bm{n})(\bm{p}_{1}\cdot\bm{S}_{1})(\bm{S}_{1}\cdot\bm{n})\nonumber
\\ & &
+\frac{3}{4}(\bm{p}_{1}\cdot\bm{n})(\bm{p}_{2}\cdot\bm{n})\bm{S}^{2}_{1}\nonumber
\\ & &
-\frac{15}{4}(\bm{p}_{1}\cdot\bm{n})(\bm{p}_{2}\cdot\bm{n})(\bm{S}_{1}\cdot\bm{n})^{2}],
\end{eqnarray}
\begin{equation}
H_{s^{2}_{2}p^{2}}=H_{S^{2}_{1}p^{2}} (1\leftrightarrow2),
\end{equation}
\begin{eqnarray}
H_{s_{1}s_{2}p^{2}}&=&\frac{\eta^{2}}{2r^{3}}\{\frac{3}{2}\{[(\bm{p}_{1}\times\bm{S}_{1})
\cdot\bm{n}][(\bm{p}_{2}\times\bm{S}_{2})\cdot\bm{n}] \nonumber \\ &
&
+6[(\bm{p}_{2}\times\bm{S}_{1})\cdot\bm{n}][(\bm{p}_{1}\times\bm{S}_{2})\cdot\bm{n}]
\nonumber \\ & & -15(\bm{S}_{1}\cdot\bm{n})(\bm{S}_{2}\cdot\bm{n})
(\bm{p}_{1}\cdot\bm{n})(\bm{p}_{2}\cdot\bm{n}) \nonumber \\ & &
-3(\bm{S}_{1}\cdot\bm{n})(\bm{S}_{2}\cdot\bm{n})(\bm{p}_{1}\cdot\bm{p}_{2})
\nonumber \\ & &
+3(\bm{S}_{1}\cdot\bm{p}_{2})(\bm{S}_{2}\cdot\bm{n})(\bm{p}_{1}\cdot\bm{n})
\nonumber \\ & &
+3(\bm{S}_{2}\cdot\bm{p}_{1})(\bm{S}_{1}\cdot\bm{n})(\bm{p}_{2}\cdot\bm{n})
\nonumber \\ & &
+3(\bm{S}_{1}\cdot\bm{p}_{1})(\bm{S}_{2}\cdot\bm{n})(\bm{p}_{2}\cdot\bm{n})
\nonumber \\ & &
+3(\bm{S}_{2}\cdot\bm{p}_{2})(\bm{S}_{1}\cdot\bm{n})(\bm{p}_{1}\cdot\bm{n})
\nonumber \\ & &
-\frac{1}{2}(\bm{S}_{1}\cdot\bm{p}_{2})(\bm{S}_{2}\cdot\bm{p}_{1})
+(\bm{S}_{1}\cdot\bm{p}_{1})(\bm{S}_{2}\cdot\bm{p}_{2}) \nonumber \\
& &
-3(\bm{S}_{1}\cdot\bm{S}_{2})(\bm{p}_{1}\cdot\bm{n})(\bm{p}_{2}\cdot\bm{n})\nonumber
\\ & &
+\frac{1}{2}(\bm{S}_{1}\cdot\bm{S}_{2})(\bm{p}_{1}\cdot\bm{p}_{2})\}
\nonumber \\ & & +\frac{3\eta^{2}}{2r^{3}\beta}\{-[(\bm{p}_{1}
\times\bm{S}_{1})\cdot\bm{n}][(\bm{p}_{1}\times\bm{S}_{2})\cdot\bm{n}]
\nonumber \\ & &
+(\bm{S}_{1}\cdot\bm{S}_{2})(\bm{p}_{1}\cdot\bm{n})^{2}\nonumber \\
& &
-(\bm{S}_{1}\cdot\bm{n})(\bm{S}_{2}\cdot\bm{p}_{1})(\bm{p}_{1}\cdot\bm{n})\}
\nonumber \\ & &
+\frac{3\eta^{2}\beta}{2r^{3}}\{-[(\bm{p}_{2}\times\bm{S}_{2})\cdot\bm{n}]
[(\bm{p}_{2}\times\bm{S}_{1})\cdot\bm{n}] \nonumber \\ & &
+(\bm{S}_{1}\cdot\bm{S}_{2})(\bm{p}_{2}\cdot\bm{n})^{2}\nonumber \\
& &
-(\bm{S}_{2}\cdot\bm{n})(\bm{S}_{1}\cdot\bm{p}_{2})(\bm{p}_{2}\cdot\bm{n})\}
\nonumber \\ & &
+\frac{6\eta}{r^{4}}[(\bm{S}_{1}\cdot\bm{S}_{2})-2(\bm{S}_{1}\cdot\bm{n})(\bm{S}_{2}\cdot\bm{n})].
\end{eqnarray}
Here, $\mathbf{p}_{1}=-\mathbf{p}_{2}=\mathbf{p}$. In a word, the
conservative Hamiltonian (1) up to the 3PN order is not completely
given until Eq. (15) appears. Clearly, Hamiltonian (1) does not
depend on any mass but the mass ratio.

The evolutions of  position $\mathbf{r}$ and momentum $\mathbf{p}$
satisfy the canonical equations of the Hamiltonian (1):
\begin{equation}
\frac{d\mathbf{r}}{dt}=\frac{\partial H}{\partial \mathbf{p}}, \quad
\quad \frac{d\mathbf{p}}{dt}=-\frac{\partial H}{\partial
\mathbf{r}}.
\end{equation}
The spin variables vary with time according to the following
relations
\begin{equation}
\frac{d\mathbf{S}_i}{dt}=\frac{\partial H}{\partial
\mathbf{S}_i}\times \mathbf{S}_i.
\end{equation}

Besides the two spin magnitudes, there are four conserved quantities
in the Hamiltonian (1), including the total energy $E=H$ and three
components of the total angular momentum vector
$\bm{J}=\bm{L}+\bm{S}$. A fifth constant of motion is absent, so the
Hamiltonian (1) is non-integrable.\footnote{Based on the idea of
[19], the Hamiltonian (1) can be expressed as a completely canonical
Hamiltonian with a 10-dimensional phase space when the canonical,
conjugate spin coordinates are used instead of the original spin
variables. If the system is integrable, at least five independent
integrals of motion beyond the constant spin magnitudes are
necessary.} Its high nonlinearity seems to imply that it is a richer
source for chaos. Next, we shall search for chaos, and particularly
investigate the effect of the NLO spin-spin interactions on the
dynamics of the system.

\section{Detection of chaos before coalescence}

With numerical simulations, we use some chaos indicators to describe
dynamical differences between the NLO spin-spin couplings excluded
and included. The appropriate ones of the indicators are selected to
study dependence of chaos on single parameter when the NLO spin-spin
couplings are included. Finally, we expect to find chaos before
coalescence by estimating the Lyapunov and inspiral decay times.

\subsection{Comparisons}

Numerical methods are convenient to study nonlinear dynamics of the
Hamiltonian (1).  Symplectic integrators are efficient numerical
tools since they have good geometric and physical properties, such
as the symplectic structure conserved and energy errors without
secular changes. However, they  cannot provide high enough
accuracies, and the computations are expensive when the mixed
symplectic integration algorithms [21,26] with a composite of  the
second-order explicit leapfrog symplectic integrator and the
second-order implicit midpoint rule are chosen. In this sense, we
would prefer to adopt an 8(9) order Runge-Kutta-Fehlberg algorithm
of variable time steps. In fact, it gives such high accuracy to the
energy error in the magnitude of about order $10^{-13}\sim10^{-12}$
when integration time reaches $10^{6}$, as shown in Fig. 1. Here,
orbit 1 we consider has initial conditions $(\bm p(0);\bm r(0))= (0,
0.39, 0; 8.55, 0, 0)$, which correspond to the initial eccentricity
$e_0=0.30$ and the initial semi-major axis $a_0=12.2$. Other
parameters and initial spin angles are respectively $\beta=0.79$,
$\chi_1=\chi_2=\chi=1.0$, $\theta_{i}=78.46^{\circ}$ and
$\phi_{i}=60^{\circ}$, where polar angles $\theta_{i}$ and azimuthal
angles $\phi_{i}$ satisfy the relations $\mathbf{\hat{S}}_{i}
=(\cos\phi_{i}\sin\theta_{i},\sin\phi_{i}\sin\theta_{i},\cos\theta_{i})$,
as commonly used in physics. The NLO spin-spin couplings are not
included in Fig. 1(a), but in Fig. 1(b). It can be seen clearly that
the inclusion of the NLO spin-spin couplings with a rather long
expression decreases only slightly the numerical accuracy.
Therefore, our numerical results are shown to be reliable although
the energy errors have secular changes.

We apply several chaos indicators to compare dynamical behaviors of
orbit 1 according to the two cases without and with the NLO
spin-spin couplings. The method of Poincar\'{e} surface of section
can provide a clear description of the structure of phase space to a
conservative system whose phase space is 4 dimensions. As a point to
note, it is not suitable for such a higher dimensional system (1).
Fortunately, power spectra, Lyapunov exponents and fast Lyapunov
indicators would work well in finding chaos regardless of the
dimensionality of phase space.

\subsubsection{Power spectrum analysis}

Power spectrum analysis reveals a distribution of various
frequencies $\omega$ of a signal $x(t)$. It is the Fourier
transformation
\begin{equation}
X(\omega)=\int_{-\infty}^{+\infty}x(t)e^{-i\omega t}dt,
\end{equation}
where $i$ is the imaginary unit. In general, the power spectra
$X(\omega)$ are discrete for periodic and quasi periodic orbits but
continuous for chaotic orbits. That is to say, the classification of
orbits can  be distinguished in terms of different features of the
spectra. On the basis of this, we know through Fig. 2 that the orbit
seems to be regular when the NLO spin-spin couplings are not
included, but chaotic when the NLO spin-spin couplings are included.
Notice that the method of power spectra is only a rough estimation
of the regularity and chaoticity of orbits. More reliable chaos
indicators are strongly desired.

\subsubsection{Lyapunov exponents}

The maximum Lyapunov exponent is used to measure the average
separation rate of
 two neighboring orbits in the phase space
and gives quantitative  analysis to the strength of chaos. Its
calculations are usually based on the variational method and the
two-particle method [27]. The former needs solving the variational
equations  as well as the equations of motion, and the latter needs
solving the equations of motion only. Considering the difficulty in
deriving the variational equations of a complicated dynamical
system, we pay attention to the application of the latter method. In
the configuration space, it is defined as [28]
\begin{eqnarray}
\lambda=\lim_{t\rightarrow\infty}\frac{1}{t}\ln\frac{|\Delta
\mathbf{r}(t)|}{|\Delta\mathbf{r}(0)|},
\end{eqnarray}
where $|\Delta\mathbf{r}(0)|$ and $|\Delta\mathbf{r}(t)|$ are the
separations between the two neighboring orbits at times 0 and $t$,
respectively. The initial distance cannot be too big or too small,
and $10^{-8}$ is regarded as to its suitable choice in the double
precision [27]. For the sake of the overflow avoided,
renormalizations from time to time are vital in the tangent space. A
bounded orbit is chaotic if its Lyapunov exponent is positive, but
regular when its Lyapunov exponent tends to zero. In this way, we
can know from Fig. 3 that orbit 1 is regular for the case without
the NLO spin-spin couplings, but chaotic for the case with the NLO
spin-spin couplings. Of course, it takes much computational cost to
distinguish between the ordered and chaotic cases.

\subsubsection{Fast Lyapunov indicators}

A quicker method to find chaos than the method of Lyapunov exponents
is a fast Lyapunov indicator (FLI). This indicator that was
originally considered to measure the expansion rate of a tangential
vector [29] does not need any renormalization, while its modified
version dealing with the use of the two-particle method [30] does.
The modified version is of the form
\begin{eqnarray}
\textrm{FLI(t)}=\log_{10}\frac{|\Delta\mathbf{r}(t)|}{|\Delta\mathbf{r}(0)|}.
\end{eqnarray}
Its computation is based on the following expression:
\begin{eqnarray}
\textrm{FLI}=-k(1+\log_{10}|\Delta\mathbf{r}(0)|)+\log_{10}
\frac{|\Delta\mathbf{r}(t)|}{|\Delta\mathbf{r}(0)|},
\end{eqnarray}
where $k$ denotes the sequential number of renormalization. The FLI
of Fig. 4(a) corresponding to Fig. 3(a) increases algebraically with
logarithmic time $\log_{10}t$, and that of Fig. 4(b) corresponding
to Fig. 3(b) does exponentially with logarithmic time. The former
indicates the character of order, but the latter, the feature of
chaos. Only when the integration time adds up to $1\times10^{5}$,
can the ordered and chaotic behaviors be identified clearly for the
use of FLI unlike the application of Lyapunov exponent. There is a
threshold value of the FLIs between order and chaos, 5. Orbits whose
FLI are larger than 5 are chaotic, whereas those whose FLIs are less
than 5 are regular.

The above numerical comparisons seem to tell us that chaos becomes
easier when the NLO spin-spin terms are included. This sounds
reasonable. As claimed in [20,21], the system (1) is integrable and
not at all chaotic when the spin-spin couplings are turned off. The
occurrence of chaos is completely due to the spin-spin couplings,
which include particularly the NLO spin-spin contributions leading
to a sharp increase in the strength of nonlinearity. In fact, we
employ FLIs to find that there are other orbits (such as orbits 2-5
in Table 1), which are not chaotic for the absence of the NLO
spin-spin couplings but for the presence of the NLO spin-spin
couplings. In addition, the strength of the chaoticity of orbits 6-8
increases. As a point to illustrate, the other initial conditions
beyond Table 1 are those of orbit 1; the starting spin unit vectors
of orbit 2 are those of orbit 1, and those of orbits 3-8 are
$\theta_{1}=84.26^{\circ}$, $\phi_{1}=60^{\circ}$,
$\theta_{2}=84.26^{\circ}$ and $\phi_{2}=45^{\circ}$. Hereafter,
only the dynamics of the complete Hamiltonian (1) with the NLO
spin-spin effects included is focused on.

\begin{table*}[thbp]
\center{ \caption{Values of FLIs and $\lambda T_d$ for different
orbits. $\textrm{FLIa}$ corresponds to the NLO spin-spin couplings
turned off. $\textrm{FLIb}$, $\lambda$, $\lambda_d$ and $\lambda
T_d$ correspond to the NLO spin-spin couplings included.}
\begin{tabular}{cccccccccccc}
\hline Orbit & $\beta$ & $\chi$ & $x$  &  $p_y$ & $e_0$ & $a_0$ &
$\textrm{FLIa}$
 & $\textrm{FLIb}$ & $\lambda$ & $\lambda_d$ & $\lambda T_d$ \\
\hline
2  & 0.90 & 1.0  & 8.55 & 0.390 & 0.30 & 12.2 & 3.6 & 10.2 & 1.9E-4 & 8.1E-4 & 0.2 \\
\hline
3  & 0.50 & 0.93 & 18.6 & 0.195 & 0.29 & 14.4 & 4.5 & 9.0  & 1.7E-4 & 3.7E-4 & 0.5 \\
\hline
4  & 0.71 & 0.95 & 17.5 & 0.200 & 0.30 & 13.5 & 3.9 & 12.4 & 2.8E-4 & 5.3E-4 & 0.5 \\
\hline
5  & 0.65 & 0.90 & 35.4 & 0.100 & 0.65 & 21.5 & 4.2 & 9.8  & 2.1E-4 & 3.8E-4 & 0.7 \\
\hline
6  & 0.86 & 0.90 & 8.40 & 0.390 & 0.28 & 11.6 & 6.5 & 20.4 & 4.2E-4 & 9.3E-4 & 0.5 \\
\hline
7  & 0.50 & 0.83 & 19.5 & 0.185 & 0.33 & 14.6 & 7.0 & 11.5 & 2.5E-4 & 3.8E-4 & 0.7 \\
\hline
8  & 0.50 & 0.97 & 18.7 & 0.190 & 0.32 & 14.1 & 9.4 & 22.5 & 5.5E-4 & 4.3E-4 & 1.3 \\
\hline
\end{tabular}}
 \label{tab1}
\end{table*}

\subsection{Lyapunov and inspiral decay times}

Taking $\beta=0.5$, the initial conditions and the initial unit spin
vectors of orbit 1 as reference, we start with the spin parameter
$\chi$ at the value 0.2 that is increased in increments of 0.01 up
to a final value of 1 and draw dependence of FLI on $\chi$ in Fig.
5(a). This makes it clear that chaos occurs when $\chi\geq0.7$.
Precisely speaking, the larger the spin magnitudes get, the stronger
the chaos gets. Note that this dependence of chaos on $\chi$ relies
typically on the choice of the initial conditions, the initial unit
spin vectors and the other parameters. As claimed in [24], there is
different dependence of chaos on $\chi$ if the choice changes. On
the other hand, taking  the initial spin angles of orbit 3, fixing
the spin parameter $\chi=0.90$ and the initial conditions $(\bm
p(0);\bm r(0))=(0, 0.39, 0; 8.4, 0, 0)$, which correspond to the
initial eccentricity $e_0=0.28$ and the initial semi-major axis
$a_0=11.6$, we study the range of the mass ratio $\beta$ beginning
at 0.5 and ending at 1 in increments of 0.01. At once, dependence of
FLI on $\beta$ can be described in Fig. 5(b). There is chaos when
$\beta\leq0.86$ and chaos seems easier for a smaller mass ratio. As
in the panel (a), this result is given only under the present
initial conditions, initial unit spin vectors and other parameters.

Do the above-mentioned chaotic behaviors occur before the merger of
the binaries? To answer it, we have to compare the Lyapunov time
$T_{\lambda}=1/\lambda$ (\textit{i.e.}  the inverse of the Lyapunov
exponent) with the inspiral decay time $T_d$, estimated by  [31]
\begin{equation}
T_{d}=\frac{12}{19}\frac{c_0^4}{\gamma}\int_{0}^{e_0}
\frac{e^{29/19}[1+(121/304)e^{2}]^{1181/2299}}{(1-e^2)^{3/2}} de,
\end{equation}
where the two parameters are
\begin{equation}
c_{0}=a_{0}(1-e^{2}_{0})e^{-12/19}_{0}(1+\frac{121}{304}e^{2}_{0})^{-870/2299}
\end{equation}
and $\gamma=64m_{1}m_{2}M/5$. When $T_{\lambda}$ is less than $T_d$
(or $\lambda T_d>1$), chaos would be observed. Because
$T_{\lambda}=3.0\times10^{3}$ and $T_d=1.3\times10^{3}$ for orbit 1,
the chaoticity can not be seen before the merger. Values of $\lambda
T_d$ for orbits 2-8 are listed in Table 1. Clearly, only chaotic
orbit 8 is what we expect. Besides these, we plot two panels (a) and
(b) of Fig. 6 regarding dependence of Lyapunov exponent on single
parameter, which correspond respectively to Figs 5(a) and 5(b). Here
are two facts. First, the results in Fig. 6 are the same as those in
Fig. 5. Second, lots of chaotic orbits whose Lyapunov times are many
times greater than the inspiral times should be ruled out, and there
are only a small quantity of desired chaotic orbits left.

In order to make the accuracy of the PN approach better, we should
choose orbits whose initial radii are larger enough than roughly
$10M$. All chaotic orbits in Table 2 are expected. Notice that the
other initial conditions of these orbits beyond this table are
$y=z=p_x=p_z=0$, and the starting spin angles are still the same as
those of orbit 3. Although an orbit has a large initial radius, it
may still be chaotic when its initial eccentricity is high enough.
This supports the result of [23] that a high eccentric orbit can
easily yield chaos.

\begin{table*}[thbp]
\center{\caption{Values of $\lambda T_d$ for chaotic orbits with big
initial radii when the NLO spin-spin contributions are included.}
\begin{tabular}{cccccccccc}
\hline
Orbit & $\chi$ & $\beta$ & $x$ & $p_y$ & $e_0$ & $a_0$ & $\lambda$ & $\lambda_d$ & $\lambda T_d$ \\
\hline
9  & 0.95 & 0.50  & 14.5  & 0.240 & 0.16 & 12.5 & 6.8E-4  & 5.2E-4 & 1.3 \\
\hline
10  & 0.97 & 0.50  & 19.5  & 0.185 & 0.33 & 14.6 & 3.8E-4  & 3.7E-4 & 1.0 \\
\hline
11  & 0.93 & 0.50  & 20.5  & 0.175 & 0.37 & 14.9 & 4.4E-4  & 3.9E-4 & 1.1 \\
\hline
12  & 0.93 & 0.50  & 25.5  & 0.140 & 0.50 & 17.0 & 4.7E-4  & 3.8E-4 & 1.2 \\
\hline
13  & 0.90 & 0.48  & 35.4  & 0.100 & 0.65 & 21.5 & 4.2E-4  & 3.5E-4 & 1.2 \\
\hline
14  & 0.90 & 0.44  & 35.4  & 0.100 & 0.65 & 21.5 & 4.3E-4  & 3.4E-4 & 1.3 \\
\hline
\end{tabular}}
 \label{tab2}
\end{table*}

\section{Conclusions}

This paper is devoted to studying the dynamics of the complete 3PN
conservative Hamiltonian of spinning compact binaries in which the
orbital part is accurate to the 3PN order and the spin-spin part as
well as the spin-orbit part includes the LO and NLO contributions.
Because of the high nonlinearity, the NLO spin-spin couplings
included  give rise to the occurrence of strong chaos in contrast
with those excluded. By scanning single parameter with the FLIs, we
obtained dependence of chaos on the parameter. It was shown
sufficiently that chaos appears easier for larger spins or smaller
mass ratios under the present considered  initial conditions,
starting unit spin vectors and other parameters. So does for a
smaller initial radius. In spite of this, an orbit with a large
initial radius is still possibly chaotic if its initial eccentricity
is high enough. Above all, there are some \textit{observable}
chaotic orbits whose initial radii are suitably large and whose
Lyapunov times are less than the corresponding inspiral times.

\begin{acknowledgements}
This research is supported by the Natural Science Foundation of
China under Grant Nos. 11173012 and 11178002.
\end{acknowledgements}

\begin{figure*}[tbph]
\center{
\includegraphics[scale=0.75]{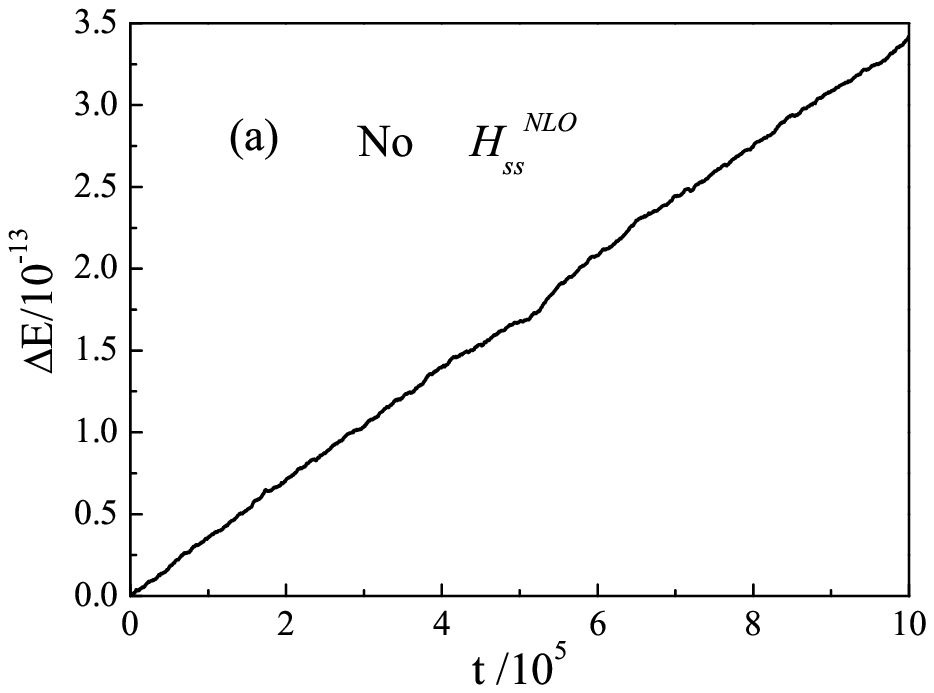}
\includegraphics[scale=0.75]{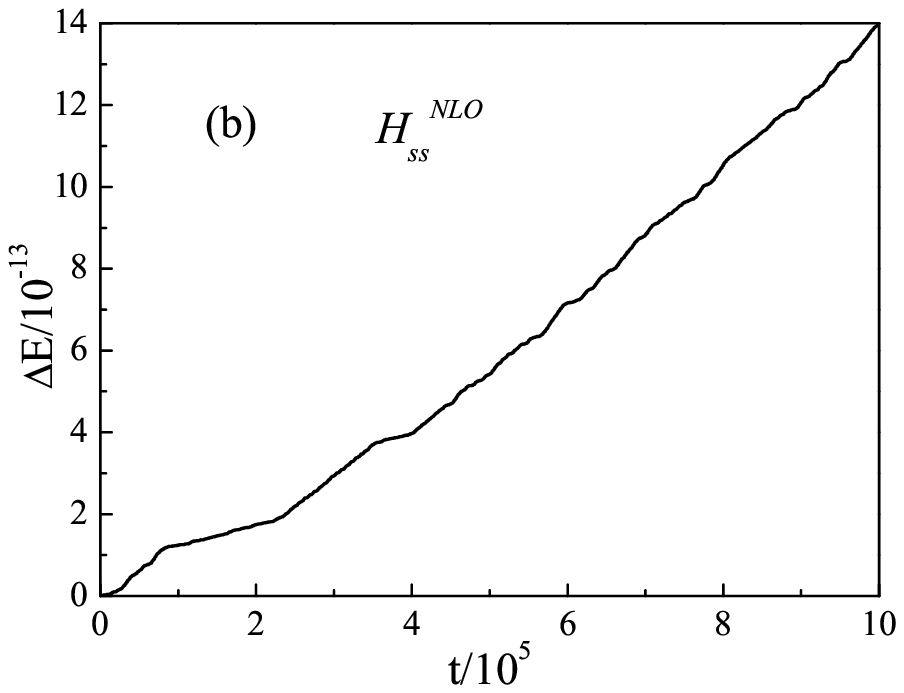}
\caption{Energy errors of orbit 1. The NLO spin-spin couplings are
not included in panel (a) but in Panel (b). }} \label{fig1}
\end{figure*}

\begin{figure*}[tbph]
\center{
\includegraphics[scale=0.75]{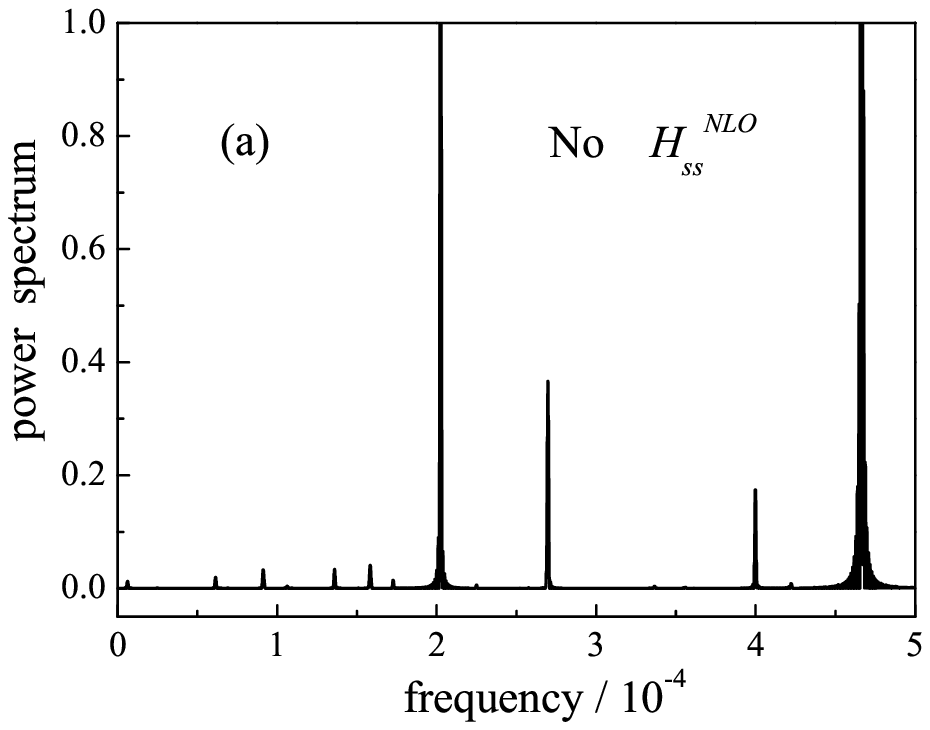}
\includegraphics[scale=0.75]{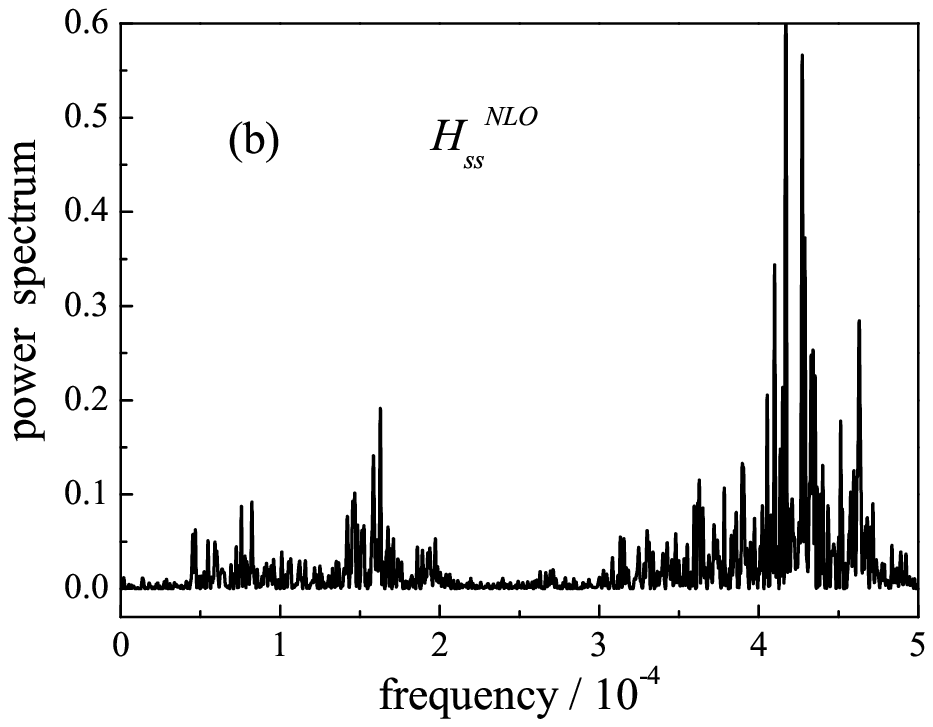}
\caption{Power spectra corresponding to Fig. 1. }} \label{fig2}
\end{figure*}

\begin{figure*}[tbph]
\center{
\includegraphics[scale=0.75]{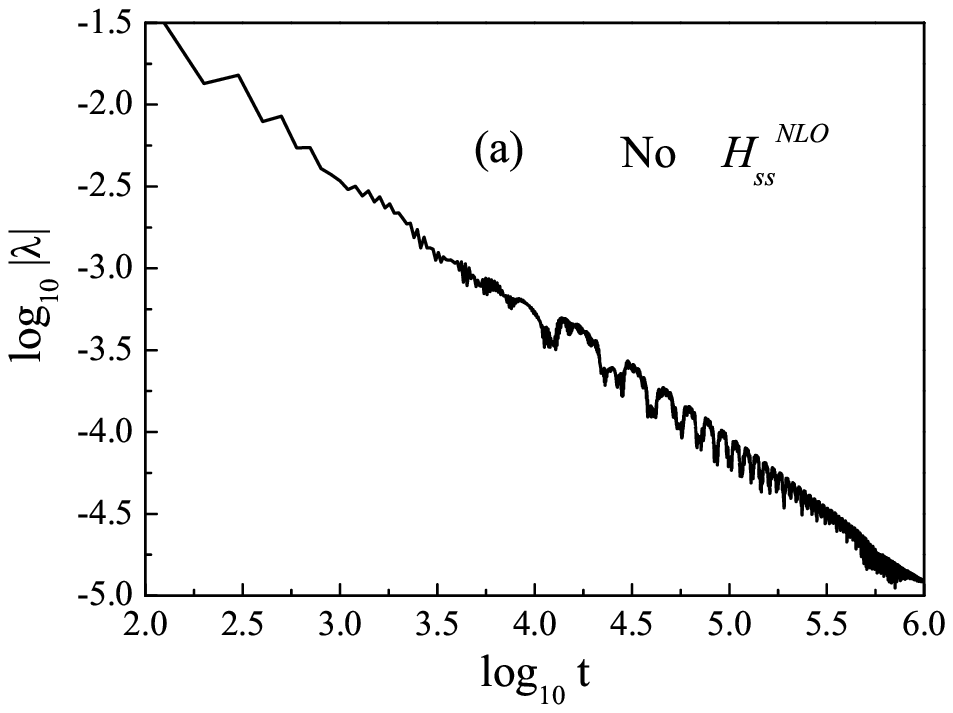}
\includegraphics[scale=0.75]{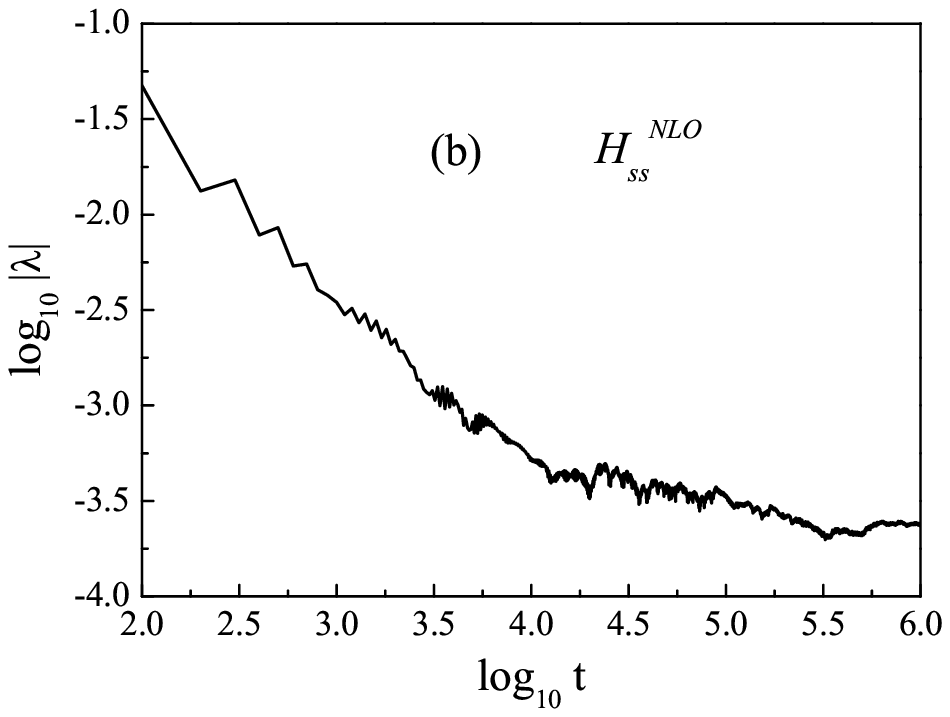}
\caption{The maximum Lyapunov exponents $\lambda$ corresponding to
Fig. 1.}} \label{fig3}
\end{figure*}

\begin{figure*}[tbph]
\center{
\includegraphics[scale=0.75]{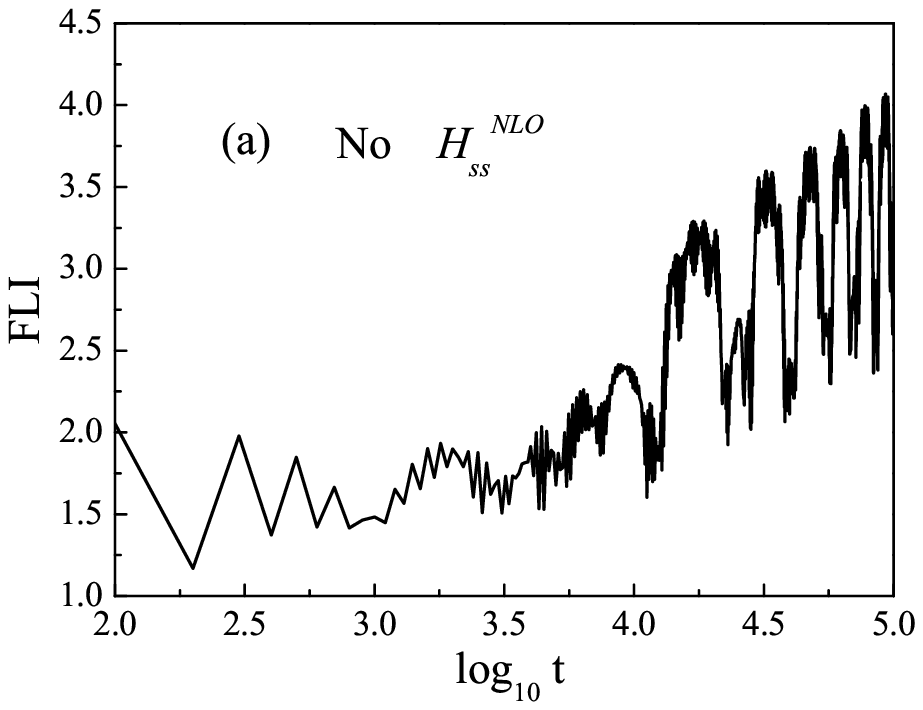}
\includegraphics[scale=0.75]{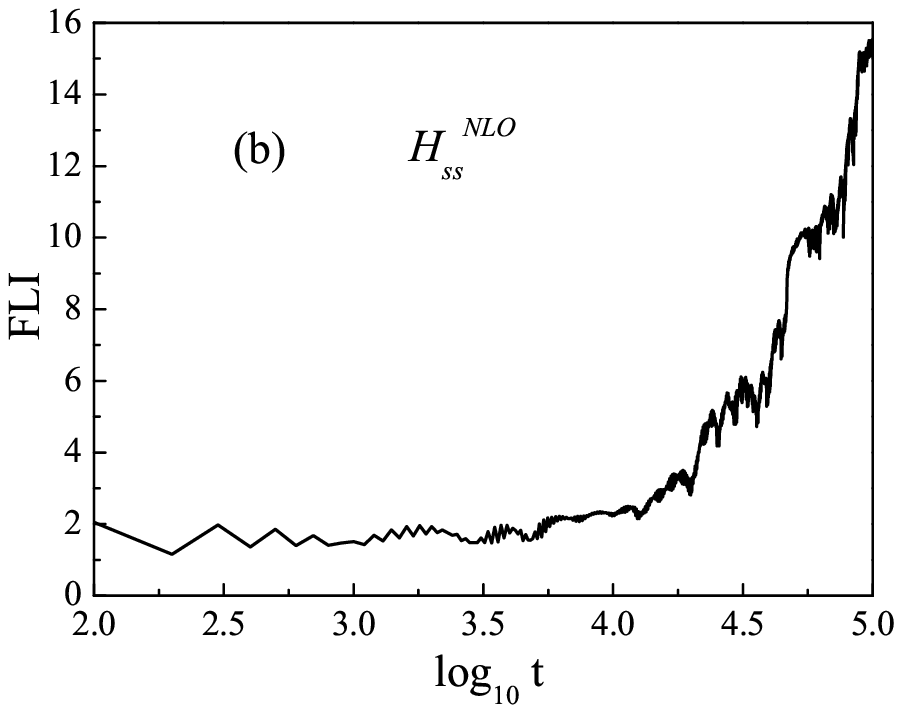}
\caption{The fast Lyapunov indicators (FLIs) corresponding to Fig.
1. }} \label{fig4}
\end{figure*}

\begin{figure*}[tbph]
\center{
\includegraphics[scale=0.75]{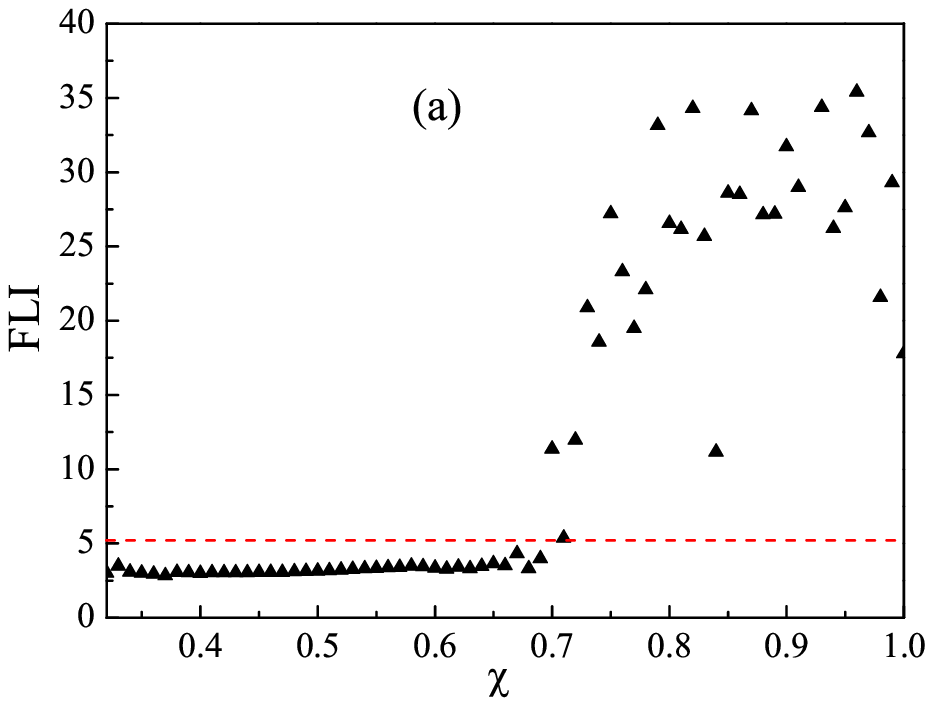}
\includegraphics[scale=0.75]{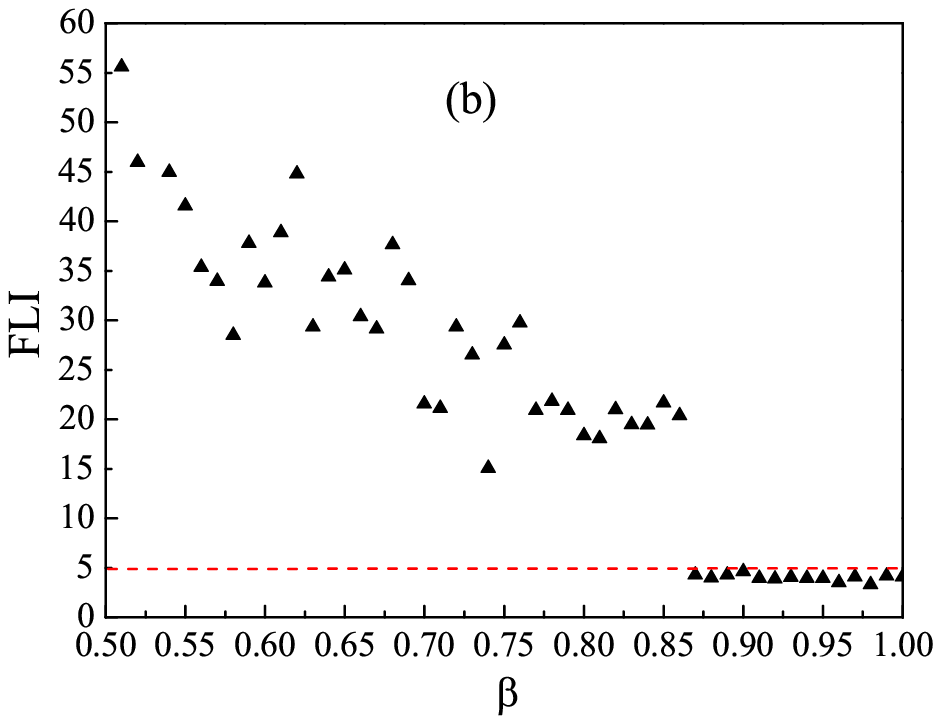}
\caption{(color online) The FLIs as a function of $\chi$ or $\beta$
when the NLO spin-spin interactions are included. All FLIs larger
than 5 mean chaos. }} \label{fig5}
\end{figure*}

\begin{figure*}[tbph]
\center{
\includegraphics[scale=0.75]{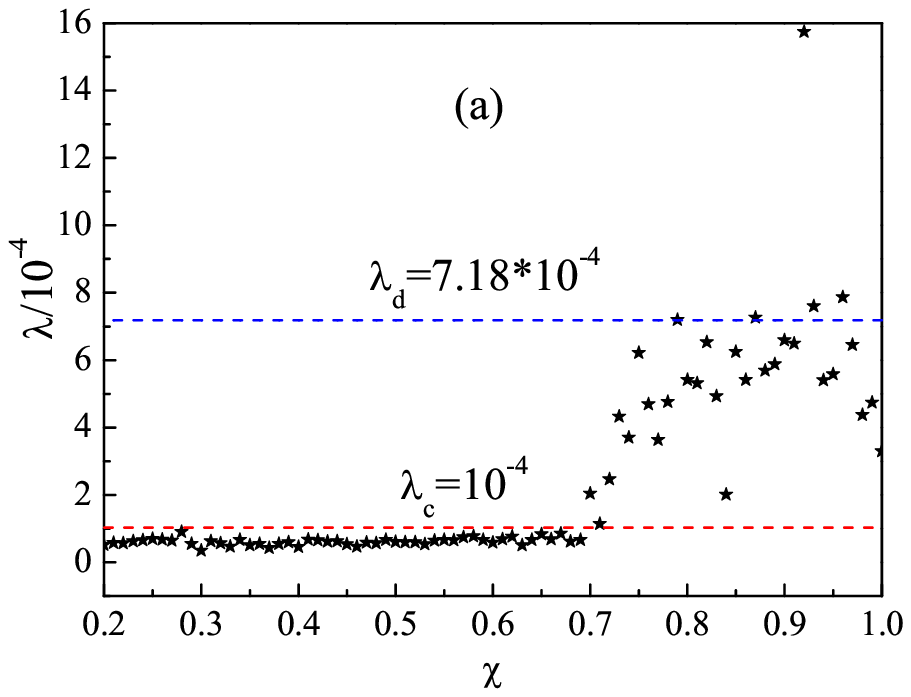}
\includegraphics[scale=0.75]{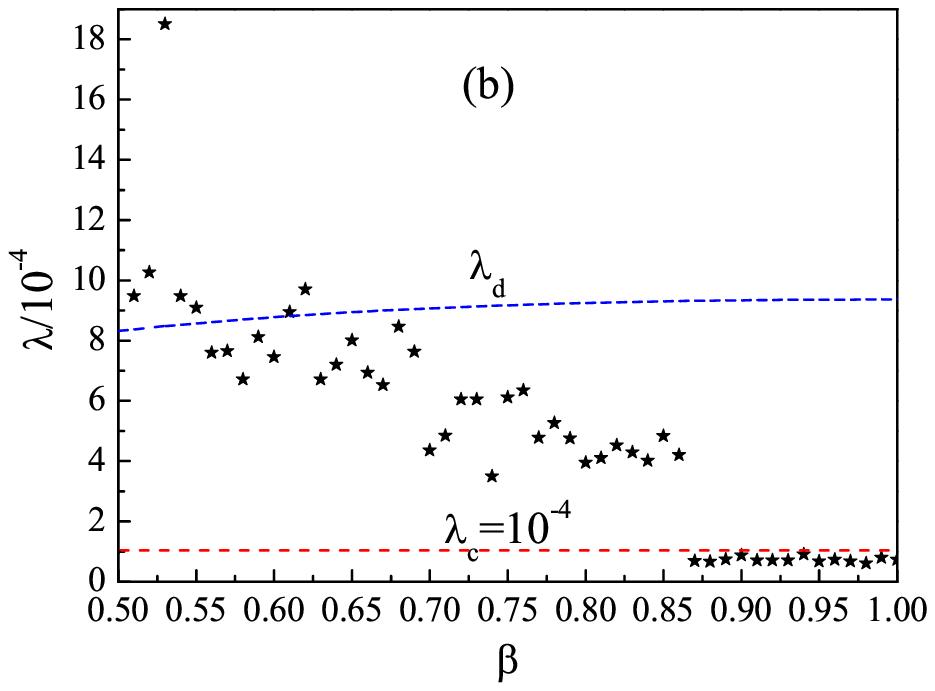}
\caption{(color online) The maximum Lyapunov exponents $\lambda$
corresponding to Fig. 5. Note that $\lambda>\lambda_c$ means chaos,
and $\lambda>\lambda_d$ with $\lambda_d=1/T_d$ indicates the
occurrence of chaos before coalescence.
 }} \label{fig6}
\end{figure*}

\end{document}